\documentstyle[12pt]{article}
\begin {document}

\begin{flushright} OITS-601\\
May 1996\\
\end{flushright}

\begin{flushleft}
\vspace{-1.5cm}
To be published in \\
Acta Physica Polonica B\\
in the special issue celebrating\\
the 60th birthday of A. Bia\l as\\
\end{flushleft}

\vskip2cm
\begin{center} {\Large \bf Beyond Intermittency:  Erraticity}
\vskip .75cm
  {\bf  Rudolph C. Hwa}
\vskip.5cm
 {Institute of Theoretical Science and Department of Physics\\  University of
Oregon, Eugene, OR 97403, USA}
\end{center}

\vskip.2cm

\begin{abstract}

Erraticity analysis of multiparticle production data is introduced as a way of
extracting the maximum amount of information on self-similar fluctuations.
It is
presented as the next logical step to take beyond the intermittency
analysis.  An
erraticity spectrum $e(\alpha)$ can be determined analogous to the multifractal
spectrum $f(\alpha)$.  An analytical example is presented to elucidate the
method of analysis and the type of results that can be obtained.
\end{abstract}

\section{A historical overview}

Andrzej Bia\l as has played an influential role in the physics of hadronic and
nuclear collisions throughout his career.  The work that he did with Robi
Peschanski on intermittency has dominated the attention of physicists working on
multiparticle production in the last ten years.  It is fitting at this
point to review
the significance of intermittency and ask where we can go from here.

When many particles are produced in high-energy collisions, the very natural
quantities to study theoretically and experimentally are averages, such as the
mean multiplicity $\left< n \right>$, the first few moments of the multiplicity
distribution $P_n$, and the rapidity distribution
$dn/dy$.  Indeed, those were the quantities investigated intensively in the
beginning of the era of multiparticle production.

Then as the collision energy was increased, the total rapidity range $Y$ became
large enough to permit meaningful partitioning of $Y$ into smaller bins of
various
sizes $\delta$.  It was found that the distributions $P_n (\delta)$ can be well
fitted by negative binomial distributions with the normalized width
increasing, as
$\delta$ decreases \cite{NA22,UA5}.  Thus began the interest in the study of
multiplicity fluctuation as a function of the bin size.  Such studies did
not catch
fire until the significance of intermittency, proposed by Bia\l as and
Peschanski
\cite{bp}, was fully appreciated.

In particle physics intermittency refers to the power-law behavior of the
normalized factorial moments $F_q$, as the bin size is decreased.  The
observation
of that behavior
\cite{ddk} therefore suggests that the mechanism for particle production has a
self-similar property.  It means that the occurrence of a large burst of
particles in
a small bin is rare, but possible if one waits long enough for such an
event to take
place.  Any model that does not possess such intermittent features is thus ruled
out.

In recent years it was found that much of the intermittency phenomenon can be
attributed to Bose-Einstein correlation among like-sign charged particles.  The
bunching of particles in small bins cannot be distinguished from the
interference
effect due to the coherent emission of same type particles from an extended
source.  While this is an important experimental finding, one should not let the
BE correlation effect completely obscure the intermittency behavior, which is
still seen in the unlike-sign charged-particle $F_2$ \cite{aga,neu}.  If it
exists in
the unlike-sign sector, then it must also exist in the like-sign sector,
though small
in comparison to the BE effect.  An effect that is small is not necessarily
unimportant.  In this case it is our only clue to an important aspect of the
dynamics of soft interaction:  self-similarity.  Hence, in my view the study of
intermittency should go on.

If the dynamics is self-similar, it is natural to ask about the multifractal
properties of multiplicity distributions.  The $G$ moments were constructed to
exhibit those properties through the multifractal spectrum $f(\alpha)$
\cite{hw1}.  The advantage is that the order $q$ of the moments $G_q$ is
continuous, and can be negative.  That makes possible the determination of
$\alpha$ which is a derivative in $q$, and facilitates the study of dips in
addition
to spikes in the rapidity distribution.  The disadvantage is that, unlike
$F_q$, the
$G_q$ moments do not filter out the statistical fluctuations automatically and
therefore require explicit elimination by ``subtraction'' \cite{der,hp}.
When that
is done, the dynamical Renyi dimensions $D_q^{dyn}$ can be determined.  For the
Monte Carlo code ECCO
\cite{hp2} that simulates hadronic collisions with intermittency, it is
found that
$D_q^{dyn}/d$ is independent of the dimension $d$ in which the $G$-moments
analysis is done
\cite{hp}.  A way to continue $F_q$ to noninteger values of $q$, while
maintaining
the attribute $F_q=1$ for all $q$ when $P_n$ is Poissonian, has been devised
\cite{hw2}.  Its application to real data has recently been attempted
\cite{Zha}.

At this point there is a slow-down in the acquisition and analysis of
multiparticle
data of hadronic collisions.  New methods of analysis have been proposed,
notably
by means of correlation integrals \cite{car} and wavelets \cite{glc}.  They are
more efficient and powerful than studying $F_q$ in discrete bins, and can
extract
more information on self-similiarity.  The application of the wavelet
analysis to
real data has not yet been done, and the reward for such an improved analysis
remains to be realized.

While the progress in phenomenology is slow, one nevertheless can ask the
theoretical question:  what is next?  Is intermittency analysis the most
that one
can do to extract information of self-similarity of the particle production
process?
In the following section a suggestion is made to carry the study to yet another
level where more information on fluctuations can be obtained.

\section{Erraticity}

Let us examine in detail the normalized factorial moments $F_q$.  The
horizontally averaged vertical moments are
\begin{eqnarray} F_q^{(v)} = \frac{1}{M} \sum_{j=1}^{M}\frac{\left<n(n-1) \cdots
(n-q+1) \right>_j}{\left< n
\right>_j^q}
\label{1}
\end{eqnarray} where $\left< \cdots \right>_j$ is the (vertical) average
over all
events of the quantity bracketed at the {\it j}th bin, $n$ being the
multiplicity in
that bin, and $M$ is the total number of bins (e.g., $M=Y/\delta$ in the
$1$-dimensional case).  If the space in which the partition into $M$ bins
is done is
made to have a flat single-particle distribution by use of the cumulative
variable
\cite{bg,och}, one can also meaningfully define the vertically averaged
horizontal
moments
\begin{eqnarray} F_q^{(h)} = \frac{1}{\cal{N}}
\sum_{e=1}^{\cal{N}}\frac{\left<n(n-1) \cdots (n-q+1)
\right>_e}{\left< n
\right>_e^q}
\label{2}
\end{eqnarray} where  $\left< \cdots \right>_e$ is now the (horizontal) average
over all bins for the {\it e}th event, and $\cal{N}$ is the total number of
events.
It is clear that the two definitions are complementary and in most
instances they
behave the same way.  In either case intermittency refers to the scaling
behavior
\begin{eqnarray} F_q \propto M^{\varphi_q}\quad,
\label{3}
\end{eqnarray} when $M$ is increased in a fixed portion of the phase space,
i.e., when the bin size
$\delta$ is decreased.

The numerator of (\ref{1}) and (\ref{2}) are nonzero only when the bin
multiplicity $n$ is $\geq q$.  Thus they pick out events and bins with large
fluctuations, $n \gg \left< n
\right>_{j,e}$, when $\delta$ is small, since $\left< n \right>_{j,e}
\propto \delta$.
It is possible that, when $q$ is large and $\delta$ is small, one may have
to wait
for many non-contributing events to go by before finding a spike that
contributes.
That is why Bia\l as and Peschanski have coined the word intermittency for the
phenomenon.  The emphasis on bin multiplicity fluctuations marked a significant
advance that intermittency generated in the subject of multiparticle production.

However, intermittency as studied so far has not fully exhausted the
characterization of fluctuations that the system can exhibit.  Let us focus on
$F_q^{(h)}$ in (\ref{2}) to be definite.  The summand is a quantity that
characterizes the ``spatial'' fluctuations (in phase space or any other
space) in an
event.  Since it plays a central role in the discussion to follow, let us
denote it by
$F_q^e$ so that
\begin{eqnarray} F_q^e = \frac{\left< n(n-1) \cdots (n-q+1) \right>_e}{\left< n
\right>_e^q}   \quad.
\label{4}
\end{eqnarray}     We then see from (\ref{2}) that $F_q^{(h)}$ is an average of
$F_q^e$ over all events.  We know that $F_q^e$ fluctuates greatly from event to
event.  Those fluctuations are ignored by the study of $F_q^{(h)}$, so
intermittency in $F_q^{(h)}$ does not fully account for all the
fluctuations that the
system exhibits.  To capture the nature of those fluctuations and to find the
associated scaling behavior constitute what can be called the erraticity
analysis,
which I now describe.

It should be remarked that the problem to be addressed is not removed by
upgrading $F_q^e$ to correlation integrals or wavelets.  We shall use $F_q^e$ as
defined in (\ref{4}) as one possible, but simple, characterization of the
spatial pattern of an event.  Other descriptions can be chosen, and can be
denoted
by $F_q^e$, used as a generic symbol.  Indeed, $F_q^e$ need not refer to
multiparticle production.  Any system that involves repeated samplings whose
outcome can fluctuate from event to event can be investigated in the erraticity
analysis.  To emphasize the generality of the method, let us simplify the symbol
$F_q^e$ to $F_e$, when the order $q$ is immaterial to the discussion.

With $F_e$ describing the spatial pattern of an event, there should exist a
distribution $P(F)$ of $F$ after many events.  Let $P(F)$ be normalized
\begin{eqnarray}
\int_{0}^{\infty} P(F) dF=1 \quad.
\label{5}
\end{eqnarray}  Clearly, $F_q^{(h)}$ in (\ref{2}) is the average
\begin{eqnarray}
\left< F \right> =\frac{1}{\cal{N}} \sum_{e=1}^{\cal{N}}F_e =
\int_{0}^{\infty} FP(F)
dF   \quad,
\label{6}
\end{eqnarray} which conveys only a small piece of the information about
$P(F)$.  Experimentally, the whole distribution $P(F)$ should be determined.
However, that may provide too much information, if the $q$ and $M$
dependences are fully explored.  Thus a few moments of $P(F)$ may be
sufficient.   Define the standard normalized moments
\begin{eqnarray}
 C_p = \frac{\left< F^p \right>}{\left< F \right>^p}   \quad,
\label{7}
\end{eqnarray} where the averages are calculated as in (\ref{6}).  The order $p$
here need not be an integer; in fact, it can even be less than $1$, but may
or may
not be less than $0$, depending on whether there are events that have
$F=0$.  For
$p>1$, $C_p$ reflects the large $F$ behavior of $P(F)$, which is sensitive
to the
spikes in phase space.  For $p<1$, $C_p$ probes the low $F$ behavior of $P(F)$,
which is influenced mainly by bins with low multiplicities, including empty
bins.
Thus knowing $C_p$ for $0<p<2$, say, reveals a great deal about the
properties of
$P(F)$, all of which are not probed by intermittency.

Collecting all the complicated properties of a complex system contributes
only to a
messy assemblage of facts.  It is only when there is some simple, universal
feature to be found in the assemblage that the phenomenological analysis
becomes worthwhile.  If there is self-similarity in the dynamics of particle
production, we should search for power-law dependence on $M$.  Eq. (3) exhibits
one such behavior.  Having generalized $\left< F \right>$ to $C_p$, it is
natural for
us to suggest the search for the scaling behavior of $C_p$
\begin{eqnarray}
 C_p \propto M^{\psi(p)} \propto \delta^{-\psi (p)}   \quad.
\label{8}
\end{eqnarray} While the behavior in (\ref{3}) has been referred to as
intermittency, we shall refer to the behavior in (\ref{8}) as erraticity.  Since
$C_p$ are the moments of $P(F)$, they describe the deviation of $F_e$ from the
mean $\left< F \right>$.  Consequently, $C_p$ is sensitive to the erratic
fluctuations
of $F_e$ from event to event.  Those fluctuations depend on the bin size because
$F_e$ itself is a description of the spatial pattern that varies according to
resolution.  Thus if those fluctuations scale with bin size, then the erraticity
exponent $\psi (p)$ is an economical way of characterizing an aspect of the
self-similar dynamics that has some order in its erratic fluctuations.

Of particular interest is an index $\mu$ defined by
\begin{eqnarray}
\mu = \left. \frac{d}{dp}\psi(p) \right|_{p=1}  \quad.
\label{9}
\end{eqnarray} It was shown in \cite{ch1,ch2} that $\mu$ is related to the
entropy in event space, and has been used to study chaotic behavior in branching
processes in QCD.

A dynamical system that has erratic fluctuations may or may not exhibit chaotic
behavior in the technical sense of chaoticity in nonlinear dynamics.  The
generalization of the notion of chaos  in classical trajectories to quantum
systems
where the degrees of freedom can increase with time is still under
investigation.
Whatever the outcome, the notion of erraticity is independent of it, and the
results  of erraticity analysis describe some features that are important
in their
own right.

For a multifractal system one usually determines the multifractal spectrum
$f(\alpha)$.  It is related to a scaling exponent $\tau (q)$ by a Legendre
transform \cite{fed}.  In multiparticle production $\tau (q)$ appears in
the scaling
law of the $G$ moments \cite{hw1}
\begin{eqnarray} G_q (\delta) \propto \delta^{\tau(q)}
\label{10}
\end{eqnarray} where $q$ is a continuous variable.  The exponent $\alpha$ is
defined by
\begin{eqnarray}
\alpha_q = \frac{d \tau (q)}{dq}
\label{11}
\end{eqnarray} and the transform is
\begin{eqnarray} f(\alpha) = q \alpha - \tau (q)   \quad.
\label{12}
\end{eqnarray} It can be shown that for a multifractal set $f(\alpha)$ is always
$\leq \alpha$, and that the information dimension $D_1$ corresponds to $D_1 =
\alpha_1 =f(\alpha_1)$ at $q=1$.

Since $C_p$ is not the same as $G_q$ (their scaling laws (\ref{8}) and
(\ref{10})
having opposite behaviors in $\delta$), our measure of erraticity does not have
multifractal properties.  Nevertheless, we still can define a spectrum
$e(\alpha)$
by Legendre transform
\begin{eqnarray} e(\alpha) = p \alpha - \psi (p)  \quad,
\label{13}
\end{eqnarray}
\begin{eqnarray}
\alpha_p = \frac{d \psi (p)}{dp}   \quad.
\label{14}
\end{eqnarray}
The function $e(\alpha)$ exhibits certain properties of erraticity more directly
than
$\psi(p)$.  For example, we have $\alpha_1 = \mu$, which is the only point where
$e(\alpha) = \alpha$.  For all other values of $\alpha$, one has $e(\alpha) >
\alpha$.  Since, by definition, $C_p =1$ at both $p=0$ and
$1$, we have
$\psi(0)=\psi (1)=0$. Thus for any $P(F)$ that becomes wider at smaller
$\delta$,
as is always the case in particle production, it follows that $\psi (p) >0$
for $p>1$,
$\psi (p) <0$ for
$0<p<1$, and
$\psi(p) >0$ for $p<0$, if $C_p$ exists.  The resultant behavior of
$e(\alpha)$ is
therefore that
$e(\alpha)=0$ at $p=0$ where $\alpha_0 <0$, and $e(\alpha) >0$ everywhere else,
where $\alpha$ is calculable.  An example of this behavior will be given in the
following section.

The values of $\alpha_0$ and $\alpha_1$ bear specific relationships to certain
averages over
$P(F)$.  Since (\ref{7}) implies
\begin{eqnarray}
\frac{d}{dp}C_p = \int_{0}^{\infty}dF\ \ell n \left(\frac{F}{\left< F
\right>}\right)\
\left(\frac{F} {\left< F \right>} \right)^p P(F) \quad,
\label{15}
\end{eqnarray} it then follows from (\ref{8}) and (\ref{14}) that in the scaling
region
\begin{eqnarray}
\alpha_0 = \frac{1}{\ell n M} \left( \left< \ell n F \right> - \ell n
\left< F \right>
\right)
\quad,
\label{16}
\end{eqnarray}
\begin{eqnarray}
\alpha_1 = \frac{1}{\ell n M} \left( \frac{\left< F\,\ell n F
\right>}{\left< F \right>}
- \ell n
\left< F \right> \right) \quad.
\label{17}
\end{eqnarray}
If the scaling laws are $\left< F \right> \propto M^\varphi$ and
$\left< \ell n F \right>
\propto \tilde{\varphi}\,\ell n M$, then we have
\begin{eqnarray}
\alpha_0 = \tilde{\varphi} -\varphi \quad,
\label{18}
\end{eqnarray}
which is negative, except for unusual $P(F)$.  On the other hand,
(\ref{17}) suggests that if we define ${\cal P} = F/\left< F \right> $, then
\begin{eqnarray}
\left< {\cal P}\,\ell n {\cal P} \right> = \alpha_1 \ell n M
\label{19}
\end{eqnarray}
in the scaling region.  The connection of $\alpha_1$ with entropy
should therefore not be surprising.

In the foregoing we have suppressed the symbol $q$ if $F$ is the normalized
factorial moment defined in (\ref{4}).  For such moments to describe the spatial
pattern, all the relevant quantities in (\ref{5}) to (\ref{19}) should be
labeled
with an index $q$, {\it viz}.\ , $F_q$,
$C_{p,q}$,
$\psi_q (p)$, $\mu_q$, $e_q(\alpha)$, $\alpha_{p,q}$, $\varphi_q$ and
$\tilde{\varphi}_q$.

\section{An analytic example}

To help make the discussion in the previous section more concrete and
transparent, let us consider an example with analytic expressions.  In real
experiments or in computer simulations the event-to-event fluctuations of $F$
may be so erratic that no simple formula can approximate
$P(F)$, let alone its dependence on $q$ and $M$.  However, there are general
trends characteristic of multiparticle production that can be built in.
Furthermore, Monte Carlo simulations of jet fragmentation in pQCD give definite
shapes of $P(F)$ that can serve as a very useful guide for the choice of
analytic
formulas.  We shall rely on the results of
\cite{ch2} to generate specific expressions.

It should first be remarked that if the bin multiplicities in an event vary
according to the Poisson distribution, then $F_q =1$ for all $q$.
Fluctuations from
$F_q=1$ have dynamical content, and usually $\left< F_q \right> >1$, for $q \geq
2$.  However, from the simulations in
\cite{ch2} we have seen that $P(F_q)$ has its maximum at $F_q =1$ for all $M$.
That is a condition that we shall impose.

We adopt the gamma distribution for $P(F)$:
\begin{eqnarray} P(F)= AF^a e^{-bF}\quad,
\label{20}
\end{eqnarray} and require that its peak be located at $F=1$.  With the
normalization (\ref{5}), it becomes
\begin{eqnarray} P(F)= \frac{a^{a+1}}{\Gamma(a+1)} F^a e^{-aF}\quad,
\label{21}
\end{eqnarray} where $a$ is the only parameter, dependent on $q$ and $M$.  All
the $F$ distributions determined in \cite{ch2} have the shapes of
(\ref{21}) with
large values of $a$.  We adopt the following parametrization to introduce
the $q$
and $M$ dependences:
\begin{eqnarray} a=500/(q\,\ell n M)^2\quad,
\label{22}
\end{eqnarray} which reproduces the general trend of the simulated results in
\cite{ch2}.

In Fig.\ 1 are shown, as examples, the distributions $P(F)$ for $q=2$ and $M=5,
50, 500$.  Clearly, for small bins (large $M$) there are large fluctuations
of $F$
from event to event.  It is those fluctuations that are measured by the moments
$C_{p,q}(M)$.  For $p=2$ the scaling behavior (\ref{8}), as shown in Fig.\ 2, is
satisfied at large $M$.  Similar scaling behaviors are found for other values of
$p$.  Thus the exponents $\psi_q (p)$ can be determined by the straightline fits
in the scaling region.  The results are shown by the dots in Fig.\ 3 for
$q=2$ and
$3$ and for some discrete values of $p$.  Those dots are well fitted by the
formula $\psi_q (p)= \sum_{i=1}^{4} b_i\,p^i$, which is then used to determine
$\alpha_{p,q}$.  It is clear from the general behavior of $\psi_q (p)$, which is
negative in the region $0<p<1$, that
$\alpha_{0,q} <0$ and $\alpha_{1,q} >0$.  Their numerical values are
\begin{eqnarray}
 \begin{array}{ll}  \alpha_{0,2} = -0.026,\quad &\alpha_{0,3} = -0.047,\\
        \alpha_{1,2} = 0.024 ,&\alpha_{1,3} = 0.04.\end{array}
\label{23}
\end{eqnarray}

With analytical formulas for $\psi_q (p)$ the spectrum $e_q(\alpha)$ can be
determined by use of (\ref{13}).  For the exponents $\psi_q (p)$ shown in
Fig.\ 3,
the corresponding $e_q(\alpha)$ are shown in Fig.\ 4.  The straightline is for
$e_q(\alpha)= \alpha$.  Thus where
$e_q(\alpha)$ curves touch the straightline are the values of
$\alpha_{1,q}$, and
where
$e_q(\alpha)=0$ give $\alpha_{0,q}$.  The entropy indices are
$\mu_q=\alpha_{1,q}$.  They get larger at higher values of $q$, which is a
consequence of the fact that $F_q$ fluctuates more from event to event at higher
$q$.

Real data are not likely to be describable by simple formulas like
(\ref{21}) and
(\ref{22}).  However, erraticity analysis can be applied to the data
directly, and
curves for $\psi_q (p)$ [using (\ref{7}) and (\ref{8})] and $e_q (\alpha)$
[using
(\ref{13}) and (\ref{14})] can be determined, if scaling behavior exists.  That
represents the ``maximum" amount of information extractable from the
horizontal and vertical fluctuations of the data that exhibit properties of
self-similarity.  Generally speaking, positive
$\alpha$ describes the spikes of the spatial distribution, while negative
$\alpha$
describes the dips.

\section{Conclusion}

Primitive averages are performed over both spatial fluctuations and event
fluctuations, which have been referred to as horizontal and vertical averages,
respectively.  Intermittency probes the scaling properties of one of those
fluctuations, and averaging over the other.  Only one kind of moments are
considered, {\it viz}., $F_q$.  Erraticity probes both types of
fluctuations, and
therefore double moments are needed:  $C_{p,q}$.

Vertical fluctuations may be due to trivial reasons, such as impact parameter
variation from event to event.  In heavy-ion collisions such variations
should be
controlled by $E_T$ cuts.  For hadronic collisions, cuts in event
multiplicity may
restrict event fluctuations too much and unduly suppress the erraticity to be
uncovered.  Those fluctuations have dynamical as well as geometrical (i.e.,
impact-parameter related) origins and should be investigated fully.  No
geometrical fluctuation is present in $e^+e^-$ annihilations, so erraticity
analysis
explores the quantum fluctuations of parton branching for every fixed
initial state
specified by the energy.

When the dynamics of particle production is known, erraticity analysis can then
describe some aspects of that dynamics, such as the (possibly) chaotic
behavior of
perturbative QCD
\cite{ch1,ch2}.  But the purpose of studying intermittency, and now
erraticity, is
to get phenomenological information from the data that can help us to learn more
about the dynamics of particle production where the theory is inoperable.
Specifically, it is for learning about the soft interaction.  Plots of
$\psi_q (p)$ or
$e_q (\alpha)$ form the arena where theory and experiment should meet.
Models of soft interaction that can reproduce the primitive averages may well
reveal deficiencies when confronted with erraticity data.

Heavy-ion data have so far not led to interesting results in intermittency
study.
Perhaps too much averaging has been done.  Erraticity analysis may reveal more
structure.

Bose-Einstein correlation has temporarily detracted the study of intermittency.
Focusing on unlike-sign charged particles in hadronic collisions and going
deeper
into erraticity may reveal features about the dynamics of soft interaction that
may finally lead to the construction of a reliable model capable of meeting all
experimental tests.

\begin{center}
\subsection*{Acknowledgment}
\end{center}

The original work on the entropy index was done in collaboration with Z.\ Cao.
This work was supported in part by the U.S. Department of Energy under Grant
No. DE-FG06-91ER40637.

\newpage
\centerline {\bf Figure Captions}

\begin{enumerate}
\item[Fig. 1{\quad}]Examples of the distribution $P(F)$ given by Eqs.\ (21) and
(22) for $q=2$.

\item[Fig. 2{\quad}]Scaling behaviors of $C_{p,q}(M)$ for $p=2$ and
$q=2,3,4$.

\item[Fig. 3{\quad}]Dependences of the erraticity exponents $\psi_q(p)$ on $p$
for $q=2,3$.

\item[Fig. 4{\quad}]The erraticity spectrum $e_q(\alpha)$ for $q=2,3$.

\end{enumerate}

\end{document}